\documentclass[pdflatex, sn-nature]{sn-jnl}

\usepackage{graphicx}%
\usepackage{multirow}%
\usepackage{amsmath,amssymb,amsfonts}%
\usepackage{amsthm}%
\usepackage{mathrsfs}%
\usepackage[title]{appendix}%
\usepackage{xcolor}%
\usepackage{textcomp}%
\usepackage{manyfoot}%
\usepackage{booktabs}%
\usepackage{algorithm}%
\usepackage{algorithmicx}%
\usepackage{algpseudocode}%
\usepackage{listings}%

\theoremstyle{thmstyleone}%

\theoremstyle{thmstyletwo}%

\theoremstyle{thmstylethree}%

\newcommand{\Rjup}{\mbox{$R_{\rm Jup}$}} 
\newcommand{\Mjup}{\mbox{$M_{\rm Jup}$}}
\newcommand{\likL}{\mathcal{L}}
\newcommand{\ltsimeq}{\raisebox{-0.6ex}{$\,\stackrel
        {\raisebox{-.2ex}{$\textstyle <$}}{\sim}\,$}}

\newcommand{\chhhh}{\mbox{${\rm CH_4}$}}
\newcommand{\hho}{\mbox{${\rm H_{2}O}$}} 
\newcommand{\coo}{\mbox{${\rm CO_{2}}$}} 
\newcommand{\nhhh}{\mbox{${\rm NH_{3}}$}} 
\newcommand{\phhh}{\mbox{${\rm PH_{3}}$}} 
\newcommand{\hhs}{\mbox{${\rm H_{2}S}$}} 
\newcommand{\hhhplus}{\mbox{${\rm H_{3}^{+}}$}}

\makeatletter
\let\jnl@style=\rm
\def\ref@jnl#1{{\jnl@style#1}}

\def\aj{\ref@jnl{AJ}}                   
\def\actaa{\ref@jnl{Acta Astron.}}      
\def\araa{\ref@jnl{ARA\&A}}             
\def\apj{\ref@jnl{ApJ}}                 
\def\apjl{\ref@jnl{ApJ}}                
\def\apjs{\ref@jnl{ApJS}}               
\def\ao{\ref@jnl{Appl.~Opt.}}           
\def\apss{\ref@jnl{Ap\&SS}}             
\def\aap{\ref@jnl{A\&A}}                
\def\aapr{\ref@jnl{A\&A~Rev.}}          
\def\aaps{\ref@jnl{A\&AS}}              
\def\azh{\ref@jnl{AZh}}                 
\def\baas{\ref@jnl{BAAS}}               
\def\bac{\ref@jnl{Bull. astr. Inst. Czechosl.}}
\def\caa{\ref@jnl{Chinese Astron. Astrophys.}}
\def\cjaa{\ref@jnl{Chinese J. Astron. Astrophys.}}
\def\icarus{\ref@jnl{Icarus}}           
\def\jcap{\ref@jnl{J. Cosmology Astropart. Phys.}}
\def\jrasc{\ref@jnl{JRASC}}             
\def\memras{\ref@jnl{MmRAS}}            
\def\mnras{\ref@jnl{MNRAS}}             
\def\na{\ref@jnl{New A}}                
\def\nar{\ref@jnl{New A Rev.}}          
\def\pra{\ref@jnl{Phys.~Rev.~A}}        
\def\prb{\ref@jnl{Phys.~Rev.~B}}        
\def\prc{\ref@jnl{Phys.~Rev.~C}}        
\def\prd{\ref@jnl{Phys.~Rev.~D}}        
\def\pre{\ref@jnl{Phys.~Rev.~E}}        
\def\prl{\ref@jnl{Phys.~Rev.~Lett.}}    
\def\pasa{\ref@jnl{PASA}}               
\def\pasp{\ref@jnl{PASP}}               
\def\pasj{\ref@jnl{PASJ}}               
\def\rmxaa{\ref@jnl{Rev. Mexicana Astron. Astrofis.}}%
\def\qjras{\ref@jnl{QJRAS}}             
\def\skytel{\ref@jnl{S\&T}}             
\def\solphys{\ref@jnl{Sol.~Phys.}}      
\def\sovast{\ref@jnl{Soviet~Ast.}}      
\def\ssr{\ref@jnl{Space~Sci.~Rev.}}     
\def\zap{\ref@jnl{ZAp}}                 
\def\nat{\ref@jnl{Nature}}              
\def\iaucirc{\ref@jnl{IAU~Circ.}}       
\def\aplett{\ref@jnl{Astrophys.~Lett.}} 
\def\apspr{\ref@jnl{Astrophys.~Space~Phys.~Res.}}
\def\bain{\ref@jnl{Bull.~Astron.~Inst.~Netherlands}} 
\def\fcp{\ref@jnl{Fund.~Cosmic~Phys.}}  
\def\gca{\ref@jnl{Geochim.~Cosmochim.~Acta}}   
\def\grl{\ref@jnl{Geophys.~Res.~Lett.}} 
\def\jcp{\ref@jnl{J.~Chem.~Phys.}}      
\def\jgr{\ref@jnl{J.~Geophys.~Res.}}    
\def\jqsrt{\ref@jnl{J.~Quant.~Spec.~Radiat.~Transf.}}
\def\memsai{\ref@jnl{Mem.~Soc.~Astron.~Italiana}}
\def\nphysa{\ref@jnl{Nucl.~Phys.~A}}   
\def\physrep{\ref@jnl{Phys.~Rep.}}   
\def\physscr{\ref@jnl{Phys.~Scr}}   
\def\planss{\ref@jnl{Planet.~Space~Sci.}}   
\def\procspie{\ref@jnl{Proc.~SPIE}}   

\begin{document}

\title[]{Methane Emission From a Cool Brown Dwarf}

\author*[1,18]{\fnm{Jacqueline K.} \sur{Faherty}}\email{jfaherty@amnh.org}
\author[2]{\fnm{Ben} \sur{Burningham}}
\author[3,4]{\fnm{Jonathan} \sur{Gagn\'e}}
\author[1]{\fnm{Genaro} \sur{Su\'arez}}
\author[1,5]{\fnm{Johanna M.} \sur{Vos}}
\author[1,8]{\fnm{Sherelyn} \sur{Alejandro Merchan}}
\author[7]{\fnm{Caroline V.} \sur{Morley}}
\author[7]{\fnm{Melanie} \sur{Rowland}}
\author[7]{\fnm{Brianna} \sur{Lacy}}
\author[9]{\fnm{Rocio} \sur{Kiman}}
\author[1]{\fnm{Dan} \sur{Caselden}}
\author[10]{\fnm{J. Davy} \sur{Kirkpatrick}}
\author[11]{\fnm{Aaron} \sur{Meisner}}
\author[12]{\fnm{Adam C.} \sur{Schneider}}
\author[13]{\fnm{Marc Jason} \sur{Kuchner}}
\author[1,14]{\fnm{Daniella Carolina} \sur{Bardalez Gagliuffi}}
\author[10]{\fnm{Charles} \sur{Beichman}}
\author[6]{\fnm{Peter} \sur{Eisenhardt}}
\author[10]{\fnm{Christopher R.} \sur{Gelino}}
\author[15]{\fnm{Ehsan} \sur{Gharib-Nezhad}}
\author[16,17]{\fnm{Eileen} \sur{Gonzales}}
\author[10]{\fnm{Federico} \sur{Marocco}}
\author[1,18]{\fnm{Austin James} \sur{Rothermich}}
\author[1]{\fnm{Niall} \sur{Whiteford}}

\affil*[1]{\orgdiv{Department of Astrophysics}, \orgname{American Museum of Natural History}, \orgaddress{\street{79th street and CPW}, \city{New York}, \postcode{10023}, \state{NY}, \country{USA}}}

\affil[2]{\orgdiv{Department of Physics, Astronomy and Mathematics}, \orgname{University of Hertfordshire}, \orgaddress{\street{AL10 9AB}, \city{Hatfield}, \country{United Kingdom}}}

\affil[3]{\orgdiv{Department}, \orgname{Plan\'etarium Rio Tinto Alcan}, \orgaddress{\street{Espace pour la Vie, 4801 av. Pierre-de Coubertin}, \city{Montr\'eal}, \country{Canada}}}

\affil[4]{\orgdiv{D\'epartement de Physique}, \orgname{Universit\'e de Montr\'eal}, \orgaddress{\street{C.P.~6128 Succ. Centre-ville QC H3C~3J7 }, \city{Montr\'eal}, \country{Canada}}}

\affil[5]{\orgdiv{School of Physics}, \orgname{Trinity College Dublin, The University of Dublin}, \orgaddress{\city{Dublin 2}, \country{Ireland}}}

\affil[6]
{\orgdiv{Jet Propulsion Laboratory},\orgname{California Institute of Technology},\city{Pasadena},\state{CA}}

\affil[7]{\orgdiv{Department of Astronomy}, \orgname{University of Texas at Austin}, \orgaddress{\street{2515 Speedway}, \city{Austin}, \postcode{78722}, \state{TX}, \country{USA}}}

\affil[8]{\orgdiv{Department of Physics \& Astronomy}, \orgname{Hunter College}, \orgaddress{\street{695 Park Ave}, \city{New York}, \postcode{10065}, \state{NY}, \country{USA}}}

\affil[9]{\orgdiv{Department of Astronomy}, \orgname{California Institute of Technology}, \orgaddress{\street{1200 E. California Blvd.}, \city{Pasadena}, \postcode{91125}, \state{CA}, \country{USA}}}

\affil[10]{\orgdiv{IPAC}, \orgname{Caltech}, \orgaddress{\street{1200 E. California Blvd.}, \city{Pasadena}, \postcode{91125}, \state{CA}, \country{USA}}}

\affil[11]{\orgname{NSF’s National Optical-Infrared Astronomy Research Laboratory}, \orgaddress{\street{950 N. Cherry Ave.}, \city{Tucson}, \postcode{85719}, \state{AZ}, \country{USA}}}

\affil[12]{\orgname{United States Naval Observatory}, \orgaddress{\street{ Flagstaff Station, West Naval Observatory Rd.}, \city{Flagstaff}, \postcode{10391 }, \state{AZ}, \country{USA}}}

\affil[13]{\orgdiv{Exoplanets and Stellar Astrophysics Laboratory}, \orgname{NASA Goddard Space Flight Center}, \orgaddress{\street{8800 Greenbelt Road}, \city{Greenbelt}, \postcode{20771 }, \state{MD}, \country{USA}}}

\affil[14]{\orgdiv{Department of Physics \& Astronomy}, \orgname{Amherst College}, \orgaddress{\street{825 East Drive}, \city{Amherst}, \postcode{01003}, \state{MA}, \country{USA}}}

\affil[15]{\orgname{NASA Ames Research Center}, \orgaddress{\street{Moffet Field}, \city{MountainView}, \postcode{94035}, \state{CA}, \country{USA}}}

\affil[16]{\orgdiv{Department of Physics}, \orgname{San Francisco State University}, \orgaddress{\street{1600 Holloway Ave}, \city{San Francisco}, \postcode{94132}, \state{CA}, \country{USA}}}

\affil[17]{\orgdiv{Department of Astronomy and Carl Sagan Institute}, \orgname{Cornell University}, \orgaddress{\street{122 Sciences Drive}, \city{Ithaca}, \postcode{14853}, \state{NY}, \country{USA}}}

\affil[18]{\orgdiv{Department of Physics}, \orgname{The Graduate Center City University of New York}, \orgaddress{\city{New York}, \postcode{10016}, \state{NY}, \country{USA}}}


\maketitle

\clearpage
\textbf{Beyond our solar system, aurorae have been inferred from radio observations of isolated brown dwarfs (e.g. \citep{Hallinan06}; \citep{Kao23}). Within our solar system, giant planets have auroral emission with signatures across the electromagnetic spectrum including infrared emission of H3+ and methane.  Isolated brown dwarfs with auroral signatures in the radio have been searched for corresponding infrared features but have only had null detections (e.g. \citep{Gibbs22}). CWISEP J193518.59-154620.3. (W1935 for short) is an isolated brown dwarf with a temperature of $\sim$482 K. Here we report JWST observations of strong methane emission from W1935 at 3.326 microns. Atmospheric modeling leads us to conclude that a temperature inversion of $\sim$300 K centered at 1-10 millibar replicates the feature. This represents an atmospheric temperature inversion for a Jupiter-like atmosphere without irradiation from a host star. A plausible explanation for the strong inversion is heating by auroral processes, although other internal and/or external dynamical processes cannot be ruled out. The best fit model rules out the contribution of H3+ emission which is prominent in solar system gas giants however this is consistent with rapid destruction of H3+ at the higher pressure where the W1935 emission originates (e.g. \citep{Helling19}). }

Brown dwarfs are a class of object that links planetary and stellar astrophysics. They have temperatures between $\sim$ 3000 -- 250~K and spectral classifications of L,T, and Y (\citep{Kirkpatrick05},\citep{Cushing11}). The Y dwarfs are a recent addition to our assortment of known compact objects and they comprise the coldest sources likely formed through the star formation process (\citep{Kirkpatrick21}).  These cold objects are directly comparable to Jupiter, with the coldest known Y dwarf -- WISE J085510.83-071442.5 -- at a temperature of $\sim$250 K (\citep{Luhman14}) -- just 100 K warmer than Jupiter (\citep{Seiff98}).  Y dwarfs present an extraordinary observational challenge for ground-based telescopes given their intrinsic faintness and need for infrared instrumentation (see e.g. \citep{Skemer16}, \citep{Faherty14}, \citep{Miles20}). JWST, a space based 6.5m infrared observatory, is perfectly suited to revolutionize our understanding of brown dwarfs and in turn exo-Jupiter atmospheres (e.g \citep{Beiler23}). In this work we report observations of two brown dwarfs obtained with JWST Cycle 1 Guest Observer (GO) program 2124.  We have obtained NIRSpec G395H spectra and mid-infrared MIRI F1000W, F1280W, and F1800W photometry for the Y dwarfs CWISEP J193518.59-154620.3  (W1935 for short) and WISE J222055.31-362817.4 (W2220 for short).  

\begin{figure}[t]
\centering
\includegraphics[width=1\textwidth]{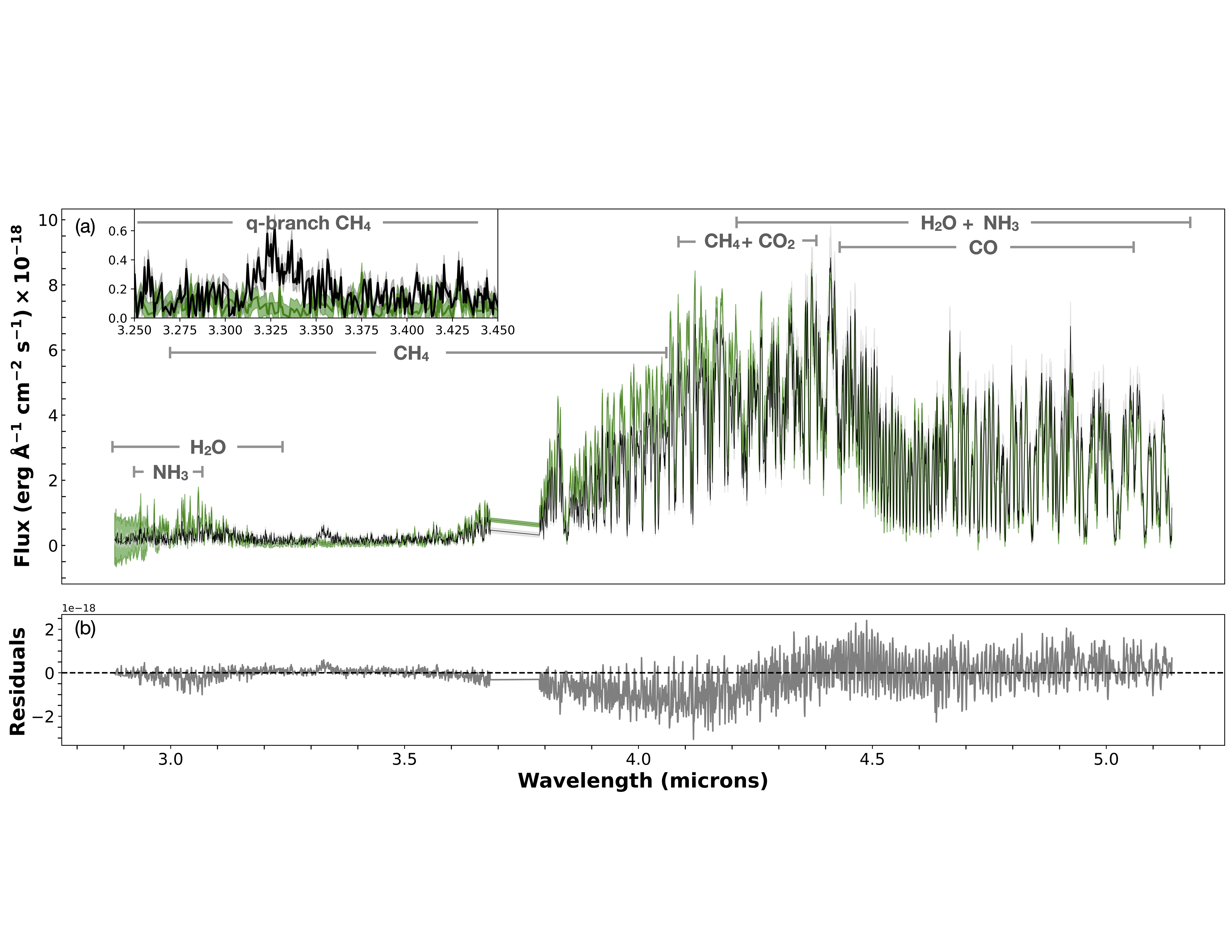} 
\caption{{\bf The JWST NIRSpec G395H spectra for the Y dwarfs W1935 and W2220.} (a) NIRSpec G395H portion of the SED for W1935 (black) compared to that of W2220 (green). Shading on both sources represents the 1$\sigma$ uncertainty on the flux.   Major opacity sources are labeled.An inset plot zooms in on the 3.326 $\mu$m CH$_{4}$ $\nu_3$ band. (b) The residuals between the two spectra.  }
\label{fig:spectra}
\end{figure}

We combined all literature data on these two objects alongside the JWST data for each source and created absolute spectral energy distributions (SEDs) which we could compare and contrast.  By integrating over the SEDs using the opensource code SEDkit (\citep{Filippazzo20}), we find that the luminosities for W2220 and W1935 are identical within uncertainties, with values of log($L_{\rm bol}$/$L_{\odot}$) equal to $-$6.4$\pm$0.1 and  $-$6.3$\pm$0.1 respectively.  Neither source has any age indications so we assumed a conservative age range of 4.5$\pm$4.0 Gyr in order to semi-empirically calculate W1935 and W2220 values of: radius 0.95$\pm$0.14 \Rjup; 0.94$\pm$0.14 \Rjup;, $T_{\rm eff}$  482$\pm$38 K; 480$\pm$41 K, $\log g$ 4.7$\pm$0.5 dex (both), and mass 6 -- 35 \Mjup\ (both). Given these objects are  indistinguishable in their fundamental parameters, they are excellent objects for inter-comparisons. 

Spectroscopically, we find that W1935 and W2220 are near clones of each other, with both showing clear and strong signatures of CH$_{4}$, CO, CO$_{2}$, H$_{2}$O, and NH$_{3}$ (see  Figure~\ref{fig:spectra}). There is one visually striking difference between the spectral characteristics of the two sources. While W2220 shows the expected CH$_{4}$ q-branch absorption centered at 3.326 $\mu$m, W1935 shows emission over that same wavelength range (see inset of 3.25 - 3.45 $\mu$m area in Figure~\ref{fig:spectra}).  

To model the 3.326 $\mu$m emission feature in W1935 as well as compare and contrast with W2220, we used the {\it Brewster} retrieval framework (\citep{Burningham17,burningham2021}) which has successfully retrieved the properties of numerous brown dwarfs (e.g. \citep{Gonzales18}, \citep{Vos22}, \citep{Calamari22} ).  For our baseline model for both objects we assumed a cloud-free atmosphere. Alongside continuum opacity sources described in detail in \citep{Burningham17,burningham2021}, we included the following gases as absorbers that would be expected to have an impact in the wavelength range covered by our data: \hho, \chhhh, CO, \coo, \nhhh, \hhs, and \phhh. 
The power of a retrieval is in its ability to parameterize the temperature-pressure (T/P) profile -- an insight into the energy distribution within a given atmosphere. {\it Brewster} can do this in a flexible way, without prescribing that the atmosphere be in radiative-convective equilibrium or have a particular slope.  This gives the T/P profile freedom to adopt whatever shape is justified by the data, including allowing an inversion where the temperature increases with altitude.  

\begin{figure}[t]
\centering
\includegraphics[width=1\textwidth]{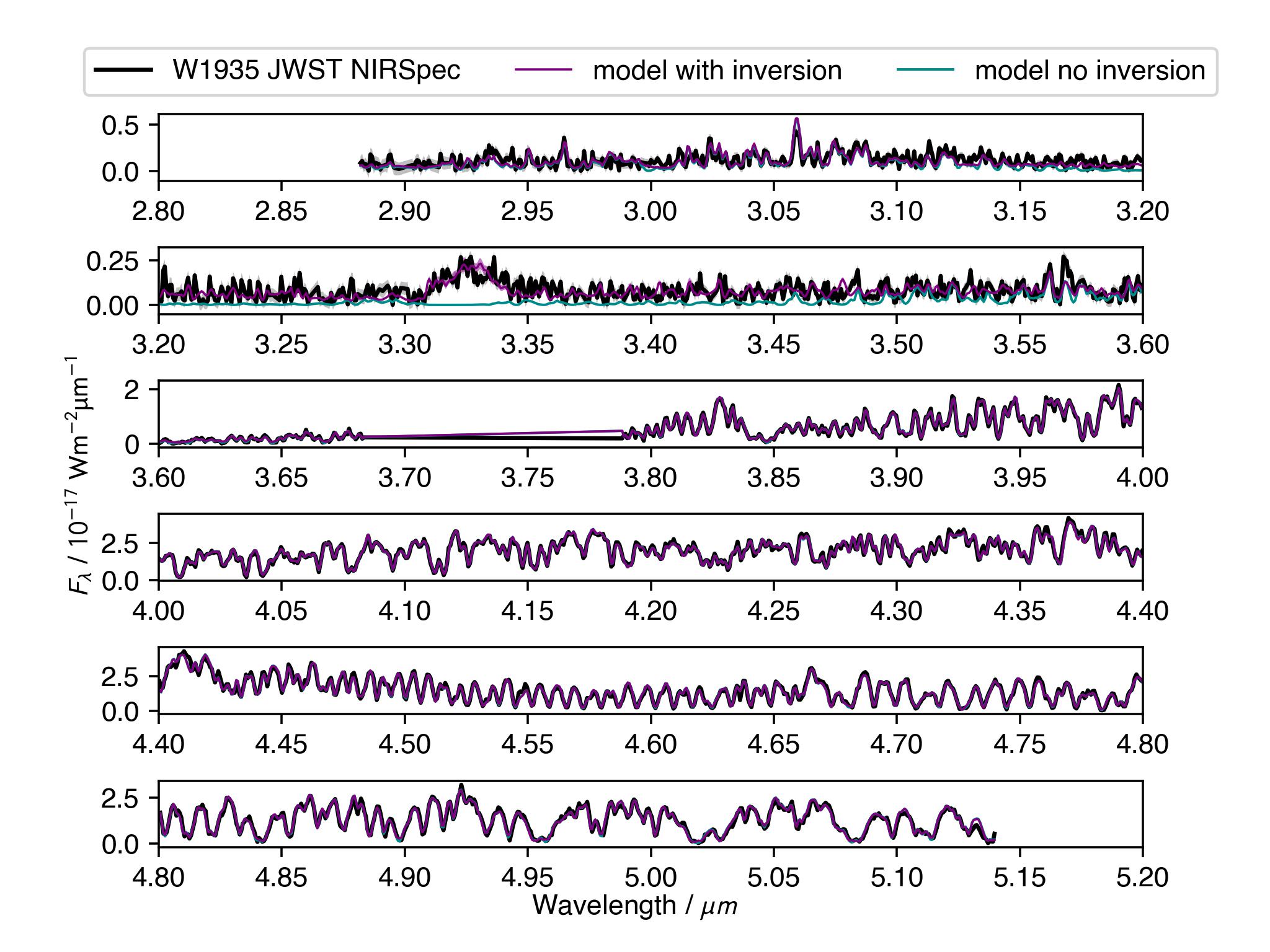} 
\caption{{\bf The JWST G395H spectrum for W1935 overlaid with the best-fit models with and without a temperature inversion.}  Overlaid in purple and dark cyan are the median retrieval models for the source with and without temperature inversions (respectively). Data uncertainties are shaded in gray. The 67\% confidence intervals in the model posteriors are shaded in dark cyan and purple, but are of comparable extent to the width of the plotted data.}
\label{fig:w1935spec}
\end{figure}

\begin{figure}[t]
\centering
\includegraphics[width=.8\textwidth]{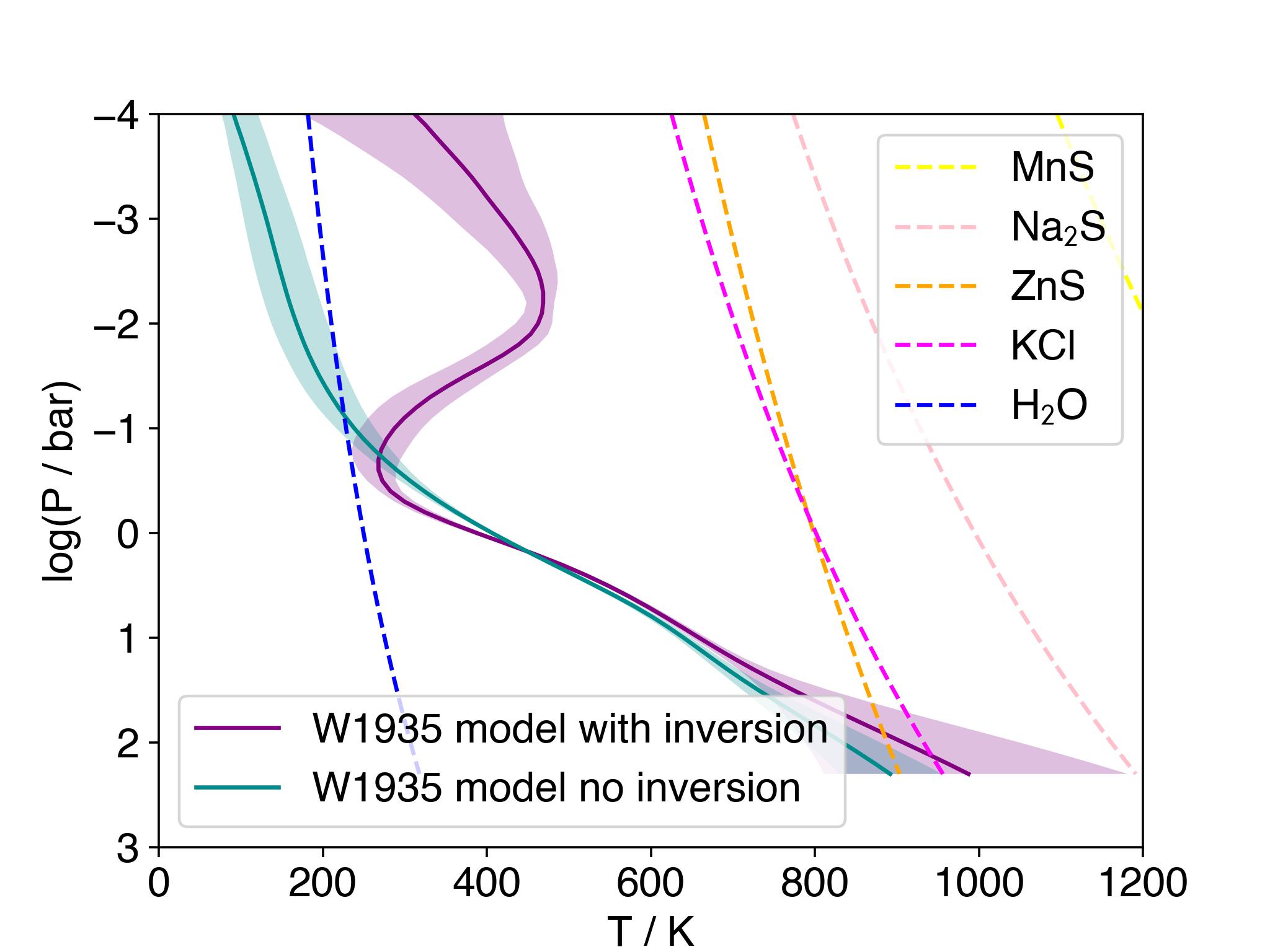} 
\caption{{\bf The retrieved thermal profiles for W1935 with and without a thermal inversion.} Median posterior profiles for the models are shown in purple and dark cyan respectively.  The 67\% confidence interval in model posteriors are indicated with shading for each. Also plotted with dashed lines are the condensation curves (assuming Solar composition gas) for possible cloud species.} 
\label{fig:retrievedprofile}
\end{figure}

The results of the retrieval verified that the two sources were near clones in all abundances (see Extended Data Table~\ref{tab:Abundances} in supplementary materials). However, the T/P profiles for the two sources show striking differences.  While the best fit retrieved T/P profile for W2220 is consistent with decreasing temperature with increasing altitude throughout the atmosphere (see Extended Data Figure~\ref{fig:w2220prof} in the methods section) as expected for an isolated source, the best fit retrieval for W1935 displays a temperature inversion of approximately 300 K centered at $\sim$1-10~millibar (see Figure~\ref{fig:retrievedprofile}). As a result, alongside our baseline model, we tested a model that forbade an inversion. For W2220, the ``no inversion" result was indistinguishable from the base result. But for W1935, the ``no inversion" retrieval could not reproduce the CH$_{4}$ emission feature (see Figure~\ref{fig:w1935spec}). 
Hence, we conclude that the observed CH$_{4}$ emission arises as a result of a thermal inversion in the stratosphere of this cool, isolated brown dwarf.

Temperature inversions have been inferred before in substellar atmospheres, both in brown dwarfs \cite[e.g.][]{lothringer2020} and giant exoplanets \cite[e.g. ][]{haynes2015}, not to mention nearly ubiquitously within the solar system. The common feature of all of these cases is the presence of an irradiating star. However, the solar system gas planets' stratospheres display temperatures even higher than can be attributed to irradiation alone \citep{strobel1973,appleby1986,MarleyMcKay1999}. Our result represents a spectacular extension of this gas giant phenomenon without any stellar irradiation. 

Much work has been dedicated to understanding the solar system cases of enhanced stratospheric heating. Both external heating by auroral processes and internal energy transport from deeper in the atmosphere by vertically propagating waves are possible mechanisms \citep[e.g. ][]{achilleos1998,bougher2005,ODonoghue2016}. The latter is a plausible explanation for the thermal inversion modeled for W1935. However, one would expect this process to be ubiquitous across a range of atmospheres. Given that this is the only non-irradiated example to date, such a universal mechanism is less likely to be responsible. 

External heating by auroral processes may be a more likely mechanism. Recent observations by \cite{odonoghue2021} have indicated that the bulk of the heating in Jupiter's upper atmosphere is driven by redistribution from hot auroral polar regions. 
In addition, alongside methane fluorescence from solar pumping, some Jovian methane emission has been tentatively attributed to heating by auroral processes \cite{kim2015}. 

Ultracool dwarfs -- a combination of the lowest mass stars and substellar mass objects -- have long been surmised to host aurora akin to those found in the giant planets of our solar system like Jupiter and Saturn. Studies have shown that $\sim$5$\%$ of ultracool dwarfs demonstrate highly circularly polarized, rotationally modulated radio pulses attributed to the electron cyclotron maser (ECM) instability - the mechanism responsible for auroral radius emission (see e.g. \citep{Hallinan06,  Kao18}). Such a low-detection rate suggests that any stratospheric heating arising from those auroral processes should be similarly rare. 

As a further test for W1935, we implemented a retrieval that included \hhhplus\-- a common emitter produced by aurorae in giant planets. Interestingly, the addition did not improve the fit for W1935 and yielded a null detection for \hhhplus\ emission. While the thermal inversion in W1935 has a similar overlying column mass to the equivalent region in the jovian atmosphere, the higher surface gravity results in a gas density that is $\sim$100 times higher. At these densities the lifetime of the \hhhplus\ ion is much shorter than its typical emission timescale \citep{Helling19}, so its absence is not surprising. 

The detection of CH$_{4}$ in emission on an object with a mass range of 6 - 35 \Mjup\ and a temperature of 482$\pm$38 K ($\sim$300K warmer than Jupiter) is enticing.  Moreover, the appearance of a temperature inversion in an object that lacks an irradiating star compounds the interest. For solar system giants with equivalent spectral emission and upper atmospheric heating, a contributing factor outside of solar pumping is auroral processes linked to nearby moons (Io for Jupiter and Enceladus for Saturn). No matter what is causing the thermal inversion and consequent methane emission on W1935, this source represents an outstanding laboratory for investigating linked phenomena that are prominent in our own solar system.

\renewcommand{\tablename}{Extended Data Table}
\renewcommand\thetable{\arabic{table}}
\setcounter{table}{0}

\renewcommand{\figurename}{Extended Data Fig.}
\renewcommand\thefigure{\arabic{figure}}
\setcounter{figure}{0}

\clearpage
\section{Methods} \label{sec:methods}

\subsection{Sample}\label{sec:sample}
\begin{figure}[h]
\centering
\includegraphics[width=1\textwidth]{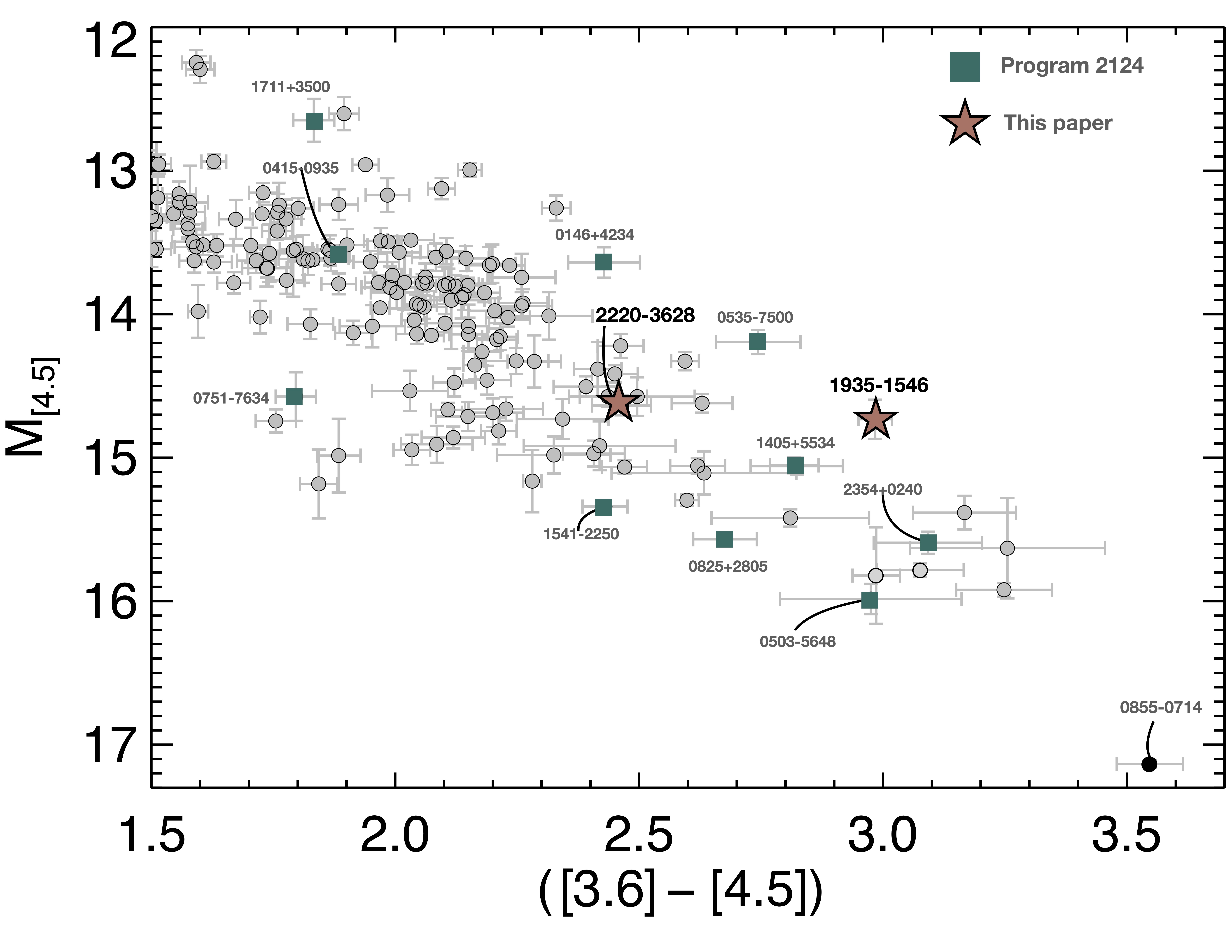} 
\caption{{\bf The Spitzer color-magnitude diagram for cold brown dwarfs.}  The sources in program 2124 are green filled squares.  Both W2220 and W1935 are highlighted by a five-point star to emphasize their position.  The 1$\sigma$ unceratinty on both color and absolute magnitude is plotted. We also label and highlight as a black filled circle the coldest known brown dwarf 0855-0714 to demonstrate the extreme of Y dwarf parameters.}
\label{fig:CMD}
\end{figure}

JWST cycle 1 GO program 2124 (PI Faherty) is obtaining G395H NIRSpec spectra for 12 cold brown dwarfs that span late-type T and Y dwarf classes.  Sources were selected by their position on the mid-infrared (Spitzer) color magnitude diagram ($M_{[4.5]}$ vs [3.6] - [4.5]) where four color bins were roughly defined to represent changing temperatures (from 800 K to $<$ 350 K).  Three sources were chosen in each color bin where one source was chosen to lie close to the median property line for the population (see for example \citep{Kirkpatrick21}), and two others were chosen from the brightest/faintest at M$_{[4.5]}$ in that color bin.  Two of the sources chosen were CWISEP J193518.59-154620.3 (W1935; \citep{Marocco19}) and WISE J222055.31-362817.4 (W2220; \citep{Kirkpatrick12}) which are the subject of this paper. At the time of the project design we speculated that the ([3.6] - [4.5]) color would define the temperature bin and we did not expect W1935 and W2220 to be so comparable. The results of this paper are strong evidence that M$_{[4.5]}$ is a better temperature binning parameter. Extended data figure~\ref{fig:CMD} shows the Spitzer color magnitude diagram for cold brown dwarfs and highlights the sample for program 2124 with the positions of W1935 and W2220 emphasized. 

\subsubsection{CWISEP J193518.59-154620.3}
CWISEP J193518.59-154620.3  (W1935 for short) was first reported in \citep{Marocco19} after a concerted effort was applied to find cold compact sources within newly analyzed Wide-field Infrared Survey Explorer (WISE) catalog data (e.g. \citep{Eisenhardt20}, \citep{Marocco21}).  The object's discovery was the result of a collaboration between the CatWISE team and the citizen science project Backyard Worlds: Planet 9 (\citep{Kuchner2017}).  Spitzer follow-up resulted in a Spitzer ([3.6] - [4.5]) color of 3.24 $\pm$ 0.31 mag, making it one of the reddest mid-infrared sources known to date.  Additional follow-up observations were done to obtain higher S/N Spitzer data in both channels and the color was refined to a Spitzer ([3.6] - [4.5]) color of 2.984 $\pm$ 0.034 mag in \citep{Meisner20}.  \citep{Kirkpatrick21} reported a parallax for this source of 69.3$\pm$3.8 mas and total proper motion of 293.4$\pm$16.3 mas yr$^{-1}$.  As noted by \citep{Meisner20} this source has a particularly low tangential velocity (vtan) compared to other Y dwarfs analyzed. The estimated spectral type from its photometry and parallax was $>$Y1 and the temperature was predicted to be 367$\pm$79K (\citep{Kirkpatrick21}).  \citep{Leggett21} obtained Gemini NIRI imaging for the source and reported MKO $J$ = 23.93$\pm$0.33 mag.

\begin{table}[t]
\begin{center}
\begin{minipage}{1.0\textwidth}
\caption{Parameters of interest for W1935 and W2220}\label{tab:target}
\begin{tabular*}{\textwidth}{@{\extracolsep{\fill}}llllll@{\extracolsep{\fill}}}
\toprule
    \bf{Astrometry} & \bf{W1935} & \bf{W2220} &\bf{units}   & \bf{ref}       \\
\midrule	
    $\alpha$ (J2000)          &    293.827684                         	&   335.231035             	& deg             	& \citep{Kirkpatrick21}, \citep{Kirkpatrick21}\\
    $\delta$ (J2000)         &     $-$15.772363                       	&  $-$36.471677             	& deg            	&  \citep{Kirkpatrick21}, \citep{Kirkpatrick21}\\
    $\varpi$          & 69.3$\pm$3.8                & 95.5$\pm$2.1               	 & mas                   &  \citep{Kirkpatrick21}, \citep{Kirkpatrick21} \\
    Distance          &   14.43$\pm$0.79            &    10.47$\pm$0.23           	& pc                	& \citep{Kirkpatrick21}, \citep{Kirkpatrick21}   \\
    $\mu_{\alpha}$    &  290.2$\pm$11.6             & 290.1$\pm$0.9     & mas yr$^{-1}$     	& \citep{Kirkpatrick21}, \citep{Kirkpatrick21} \\
    $\mu_{\delta}$    &   43.1$\pm$11.5             & -97.1$\pm$0.9     & mas yr$^{-1}$     	& \citep{Kirkpatrick21}, \citep{Kirkpatrick21} \\
\toprule
    \bf{Kinematics}        &
            &
            &
       &
 	       \\
\midrule	
   X   & 	12.69 $\pm$0.69  &	5.67 $\pm$ 0.12			& pc		& This paper\\
   Y   & 	5.49$\pm$  0.30  & 	0.70 $\pm$  0.02		& pc		& This paper \\
   Z   & 	-4.14 $\pm$ 0.22 & -8.77 $\pm$ 0.19			& pc		& This paper\\
   U   &	-41.20$\pm$ 4.52 &  -39.61$\pm$1.54			& km\,s$^{-1}$		& This paper\\
   V   &   -6.03$\pm$2.12    & -11.10$\pm$0.25			& km\,s$^{-1}$		& This paper\\
   W   &	-5.61$\pm$ 1.87 &  37.00$\pm$2.35			& km\,s$^{-1}$		& This paper\\
   RV 	&	-36.9$\pm$5.1		 &	-53.2$\pm$2.8		& km\,s$^{-1}$		& This paper\\
   Total Velocity & 42.02$\pm$5.33 & 55.33$\pm$2.82& km\,s$^{-1}$		& This paper\\
\toprule
    \bf{Photometry}        &
            &
            &
       &
 	       \\
\midrule	
    SpT                 &$>$Y1         &     Y0                &               & \citep{Kirkpatrick21}, \citep{Schneider15} \\
    MKO $Y$				&$\cdots$			&20.91$\pm$0.09			& mag			& \citep{Leggett16}\\
    MKO $J$				&23.93$\pm$0.33	& 20.64$\pm$0.05		& mag			& \citep{Leggett21}, \citep{Leggett16} \\
    MKO $H$				&$\cdots$			&20.96$\pm$0.08			& mag			&\citep{Leggett16} \\
    MKO $K$				&$\cdots$			&21.33$\pm$0.15			& mag			& \citep{Leggett16}\\
    WISE W1				&18.75$\pm$0.30	& 18.29$\pm$0.15 			& mag			& \citep{Kirkpatrick21}, \citep{Kirkpatrick21}\\
    WISE W2				&15.79$\pm$0.06	& 14.81$\pm$0.02			& mag			& \citep{Kirkpatrick21}, \citep{Kirkpatrick21}\\
    Spitzer [3.6]					&18.51$\pm$0.03	& 17.20$\pm$0.06 			& mag			& \citep{Meisner20}, \citep{Kirkpatrick21} \\
   Spitzer [4.5]					&15.53$\pm$0.02	& 14.74$\pm$0.02			& mag			& \citep{Meisner20}, \citep{Kirkpatrick21} \\
    JWST F1000W	& 13.740$\pm$0.005				&13.354$\pm$0.006			& mag			& This study \\
    JWST F1280W		&	13.126$\pm$0.007			&	12.979$\pm$0.009		& mag			& This study \\
    JWST F1800W 				&	12.107$\pm$0.017			&	12.276$\pm$0.021		& mag			& This study \\
\toprule
    \bf{Fundamental Parameters}        &
            &
            &
       &
 	       \\
\midrule	
   Age					          &	4.5$\pm$4.0		&	4.5$\pm$4.0		& Gyr			& This study \\
   $T_{\rm eff}$                     & 482$\pm$38      &	480$\pm$41 		& K & This study \\
   log($L_{\rm bol}$/$L_{\odot}$)&	-6.3$\pm$0.1	&	-6.4$\pm$0.1 		&  & This study \\
 Radius    				      &	0.95$\pm$0.14	&	0.94$\pm$0.14		& \Rjup\	& This study \\
   Mass       				      &	6 -- 35			&	6 -- 35			& \Mjup\	& This study \\
   $\log g$				      &	4.7$\pm$0.5	&	4.7$\pm$0.5		& dex		& This study \\
\botrule
\end{tabular*}
\end{minipage}
\end{center}

\end{table}

\begin{table}[t]
\begin{center}
\begin{minipage}{\textwidth}
\caption{Retrieved Gas Abundances and Fundamental Parameters}\label{tab:Abundances}
\begin{tabular*}{\textwidth}{@{\extracolsep{\fill}}lccccc@{\extracolsep{\fill}}}
\toprule%
\bf{$\log f_{gas}$}                                 & 
\bf{W1935} & 
\bf{W2220} \\
\midrule
H$_{2}$O & $-2.69^{+0.07}_{-0.11}$ & $-2.60^{+0.10}_{-0.17}$\\
CH$_{4}$ & $-2.81^{+0.07}_{-0.11}$ & $-2.79^{+0.10}_{-0.17}$ \\
CO$_{2}$ &$-7.65^{+0.07}_{-0.11}$  & $-7.55^{+0.10}_{-0.17}$\\
CO & $-5.09^{+0.08}_{-0.11}$  & $-4.92^{+0.10}_{-0.17}$\\
NH$_{3}$ &$-3.97^{+0.08}_{-0.11}$  & $-4.01^{+0.10}_{-0.17}$ \\
H$_{2}$S &$-4.19^{+0.09}_{-0.12}$  & $-4.27^{+0.14}_{-0.18}$\\
PH$_{3}$ & $\ltsimeq -9$  & $\ltsimeq -9$\\

\hline
\hline
\toprule%
\bf{Metallicity and C/O}\\
\midrule
$[M/H]$ & $0.45^{+0.07}_{-0.12}$ & $0.51^{+0.11}_{-0.18}$ \\
C/O & $0.77^{+0.04}_{-0.03}$ & $0.64^{+0.03}_{-0.03}$ \\
\botrule

\end{tabular*}
\end{minipage}
\end{center}
\end{table}

\subsubsection{WISE J222055.31-362817.4}
WISE J222055.31-362817.4 (W2220 for short) was first reported in \citep{Kirkpatrick12} after a search of the WISE catalog (\citep{Wright10}) for cold compact objects.  In \citep{Kirkpatrick12} it was followed up with the Spitzer Space Telescope, Keck/NIRSpec-N3 and Keck/NIRSpec-N5 as well as AAT/IRIS2 and SOAR/OSIRIS. The source was confirmed as a cold brown dwarf with WISE W1, W2, Spitzer [3.6], [4.5], and MKO $J$ $H$ photometry as well as NIR spectra.  \citep{Schneider15} obtained an HST grism spectrum for the source which confirmed the Y0 classification with higher S/N data.  Follow-up photometric and astrometric imaging was done using both space-based (e.g. Spitzer) and ground-based (e.g. FourStar at Magellan) instruments and the most recent/highest S/N trigonometric parallax is 95.5$\pm$2.1 mas (see \citep{Tinney14}, \citep{Beichman14}, \citep{Martin18}, \citep{Kirkpatrick19}, \citep{Kirkpatrick21} for astrometric history of the source).  \citep{Kirkpatrick21} estimated a temperature of 452$\pm$88 K for this source based on its position on the color-magnitude diagram.    

\subsection{The Data} \label{sec:data}
JWST Program 2124 obtained both NIRSpec G395H spectra and MIRI F1000W, F1280W, and F1800W photometry to fill out the peak of the spectral energy distribution (SED) and the Rayleigh-Jeans tail of the SED for 12 brown dwarfs. NIRSpec data was obtained using the F290LP filter, the G395H grating, the S200A1 aperture and the SUB2048 subarray.  The resultant wavelength coverage ranged from 2.87 to 5.14 $\mu$m with a resolving power of $\sim$2700.  Acquisition images were first obtained for each target using the WATA method, the CLEAR filter, and the NRSRAPID readout pattern.  W2220 was observed with NIRSpec on 4 November 2022 with 28 groups per integration, 10 integrations per exposure and 3 total dithers for a summation of 30 total integrations in 1356 seconds of exposure time.  Recorded time including overhead for the W2220 NIRSpec observation was 1.42 hours.   W1935 was observed with NIRSpec on 17 October 2022 with 46 groups per integration, 11 integrations per exposure and 3 total dithers for a summation of 33 total integrations in 2417 seconds of exposure time. Recorded time including overhead for the W1935 NIRSpec observation was 1.76 hours.

MIRI photometry was obtained with the F1000W, F1280W, and F1800W filters.  For each filter the FASTR1 readout pattern was chosen with a 2-point dither pattern.  W1935 was observed with MIRI on 20 September 2022 using 15 groups per integration for F1000W, 13 groups per integration for F1280W, and 11 groups per integration for F1800W.  Total exposure time plus overhead for the MIRI observing of W1935 was 1.03 hours.  W2220 was observed with MIRI on 18 October 2022 using 7 groups per integration for F1000W, 7 groups per integration for F1280W, and 10 groups per integration for F1800W.  Total exposure time plus overhead for the MIRI observation of W2220 was 0.54 hours. 

As noted in Section \ref{sec:sample} above, both W2220 and W1935 have previously published photometry and W2220 has previously published spectra.  Extended Data Table ~\ref{tab:target} lists all observables both previous and new within this paper, included in the analysis that follows. 

\subsection{Data reduction} \label{sec:datareduction}
We used the official JWST science calibration pipeline (version 1.10.0) to reduce the NIRSpec G395H spectra starting from uncalibrated data downloaded from the Mikulski Archive for Space Telescopes (MAST). The pipeline is comprised of three separate stages. Stage 1 of the pipeline performs detector-level corrections (e.g. bias subtraction, dark subtraction, and cosmic-ray detection) and ramp fitting to generate a count rate image for the individual uncalibrated image of each exposure. The resulting count rate images were calibrated by applying instrument-level and observing-mode corrections in Stage 2. Stage 3 combines multiple calibrated exposures and extracts the spectrum. 
We optimized the aperture extraction location by using the relative slit position of the target to account for inaccuracies in the object coordinates and/or celestial WCS.

The flux uncertainties automatically propagated through the pipeline were all null for the extracted spectrum due to the most recent reference flat files for NIRSpec having NaN values. We recalculate flux errors for the extracted spectrum by combining in quadrature the Poisson variance (\texttt{FLUX\_VAR\_POISSON}) and read noise variance (\texttt{FLUX\_VAR\_RNOISE}) provided in the extracted spectrum file. 

Looking at the W1935 3.326 $\mu$m spectral feature alone, we carefully examined each dither position to confirm that the feature was present.  While this is in general a low signal to noise area of the spectrum (average of 5 - 10 across the whole feature) we find the methane emmision persists in individual single exposure dithers confirming its presence.

For both W2220 and W1935 we used the MAST produced pipeline reductions of MIRI photometry choosing the aper total vegamag column as our preferred magnitude.

\subsection{Radial Velocity} \label{sec:RV}
Given the high resolution of G395H data, we were able to compute radial velocities for both targets. All values reported are from a correlation with the \citep{Lacy23} models. Our values for W1935 and W2220 which are listed in Extended Data Table ~\ref{tab:target}, came out to be -36.9$\pm$5.1 km s$^{-1}$ and -53.2$\pm$2.8 kms$^{-1}$ respectively.  We used these values in the banyan sigma kinematic code \citep{Gagne18} to check if our targets belonged to any known moving group. For both objects the full kinematics yielded a 99.9\% field population probability.

We also examined the total space velocity for each object and computed values of 42.02$\pm$5.33 km s$^{-1}$ and 55.33$\pm$2.82 km s$^{-1}$ for W1935 and W2220 respectively. These values are consistent with the field brown dwarf population (see for instance \citep{Faherty09}).  Consequently we have no evidence for youth or significant old age in the observables. Therefore, we choose a broad age range of 4.5$\pm$4.0 Gyr to estimate fundamental parameters in the SED approach (see below).  

\subsection{Spectral Energy Distribution Construction and Results} \label{sec:SED}
Key to the analysis contained within this work is generating distance calibrated spectral energy distributions (SEDs) inclusive of the newly obtained JWST data and all previously collected observables.  To construct the SEDs, we used the open-source package SEDkit (\citep{Filippazzo20}) first published and used by \citep{Filippazzo15} to analyze the fundamental parameters of brown dwarfs.  As described in \citep{Filippazzo15}, we first combine the parallax with spectra and photometry.  For W1935, the only spectrum available was the G395H data.  For W2220 there was an HST grism spectrum available from \citep{Schneider15} as well as the newly acquired JWST data. All photometry used in the SED for both objects is listed in Extended Data Table~\ref{tab:target}.  

Using SEDkit and the input data, we constructed the SEDs, integrated under the data as described in \citep{Filippazzo15} and calculated the $L_{bol}$ value for each object. In order to examine the similarities and differences across the two objects, Figure~\ref{fig:spectra} shows the output SEDs for both W2220 and W1935 scaled to 10pc. As described in the main text, the age estimation was paired with the \citep{Marley21} evolutionary models to acquire a radius range and then semi-empirically calculate the 
$T_{\rm eff}$, $\log g$, and mass. All values are listed in Extended Data Table ~\ref{tab:target}.  

Figure~\ref{fig:spectra} shows the G395H portion of the SEDs overplotted.  The SEDs are both excellent fits for each other but show important differences.  The 3.326 $\mu$m CH$_{4}$ band is striking given that it is clearly in emission for W1935 but absent for W2220.  There is also a difference in the 3.8 -- 4.3 $\mu$m region, where W2220 appears to have the same molecular features (see for instance Figures ~\ref{fig:w1935spec} - ~\ref{fig:w2220spec}) but the gas is warmer hence flux is higher for W2220 driving its bluer MIR color. 

\subsection{Retrieval Analysis}
\label{sec:brewster}
We carried out retrieval analysis of the NIRSpec spectroscopy of our two targets using the {\it Brewster} package. {\it Brewster} is publicly available and was developed to model substellar atmospheres. It has been successfully applied to a range of brown dwarf atmospheres from cool T dwarfs to hot L type subdwarfs, including cloudy and planetary mass objects \citep{Burningham17,Gonzales18,Gonzales20,burningham2021,Calamari22,Gaarn2022,Vos22}. This is its first application in the Y~dwarf regime. 

\subsubsection{Retrieval method}

{\it Brewster} consists of a forward model and sampler. The forward model produces a synthetic spectrum for a given set of atmospheric parameters. This is coupled to a Bayesian sampler which explores the parameter space and estimates the posterior distribution for the forward model parameters given the data.  

The forward model is a 1D radiative transfer layered atmosphere model consisting of 64 layers distributed uniformly in $\log$-pressure (in bars) space between 2.3 and -4.0. 
The model evaluates the emergent flux using the two stream source function technique of \cite{toon1989}, including scattering \citep{mckay1989}.  
This requires each layer in the atmosphere to be specified in terms of wavelength dependent optical depth, single scattering albedo, and asymmetry parameter, as well as the temperature of the gas in the layer. 
The resultant spectrum is then processed to allow direct comparison to the data and the calculation of a likelihood for the fit.

The parameters for the forward model fall into the following groups: gas-phase opacity, cloud opacity, temperature structure, global properties of the target. 
The gas-phase phase opacity is set by the choice of absorbing gases, their concentration and distribution in the model atmosphere. 
We included the following gases in this study: \hho, \chhhh, CO, \coo, \nhhh, \hhs, and \phhh. 
In this case we parameterized the concentrations 
 of the absorbing gases as vertically constant volume mixing ratios as in previous studies with this code. 
 This approach provides substantial flexibility in arriving at possible solutions while still retaining computational simplicity. 
 We neglected cloud opacity in this study and do not discuss this aspect of {\it Brewster} any further here but leave that work for a future study. 

 We modelled the temperature structure using the same method as \cite{Line2015,Line17}. Briefly, we parameterize the temperature at 13 evenly spaced points in log-pressure, and evaluate the temperature in our atmospheric layers using spline interpolation from these. 
 We avoid implausible discontinuities and so-called ``ringing" by penalizing the second derivative of the thermal profile in the likelihood by using the following log-prior on the temperature:

 \begin{equation}
     \ln{p({\rm \mathbf{T}})} = -\frac{1}{2\gamma} \sum^N_{i = 1}(T_{i+1} -2T_i +T_{i-1})^2 - \frac{1}{2}\ln{(2\pi\gamma)}
 \end{equation}
 
This sums the discrete second derivative of the temperature $T$ at each level $i$, and weights it by $\gamma$. The parameter $\gamma$ sets the degree to which the likelihood will be penalized by wiggles in the thermal profile, and is included in our retrieval as a free parameter. By including this as a free-parameter, the data can set the degree of smoothing imposed on the profile. 
We follow \cite{Line2015} in adopting an inverse $\Gamma$ distribution as the hyperprior for $\gamma$, with properties specified in Extended Data Table~\ref{tab:priors}.

We parameterize the global properties of the target as follows. The radius of the source is encoded in the scale factor that is applied to the top-of-atmosphere flux produced by our radiative transfer code to allow comparison to the flux received by NIRSpec. This scale factor is equal to $R^2 / D^2$, where $R$ is the radius of the source, and $D$ is its heliocentric distance.  
Our forward model uses $R^2 / D^2$, and hence this is the quantity that is estimated directly by the retrieval. 
We use our knowledge of the target's distance to estimate the radius post hoc, and incorporate the uncertainties in both the distance and the absolute flux calibration of the spectrum in the error estimate for the radius. 
The surface gravity for the source is parameterized as $\log g$, where $g$ is the gravitational acceleration at the altitude of our model atmosphere in ${\rm cm s^{-2}}$.
We combine this parameter with our post hoc estimate for the radius to estimate the mass of the source. 
Finally we include parameters to apply rotational broadening and radial velocity shifts to our model spectrum. 
We use the rotational broadening code provided by \cite{rotbroad} to achieve the former. 

\begin{figure}[h]
\centering
\includegraphics[width=1.0\textwidth]{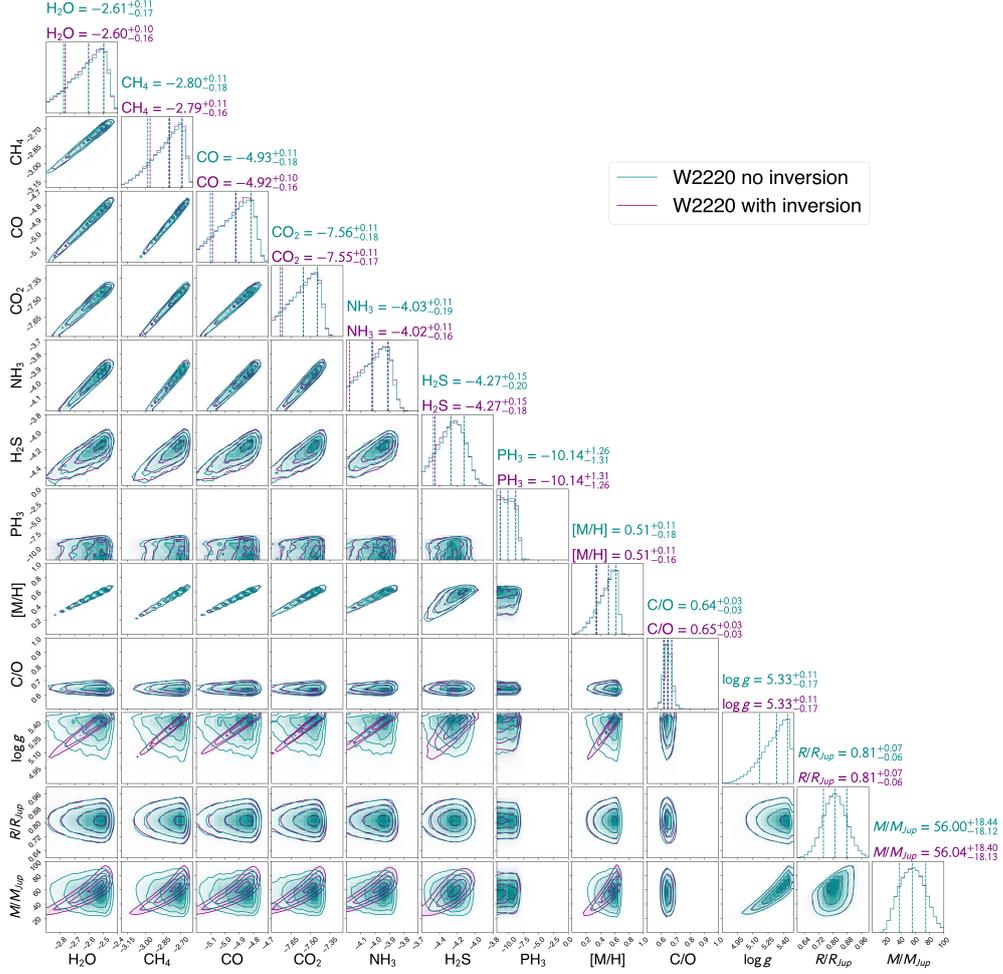} 
\caption{{\bf The retrieved composition of W2220, along with gravity, and derived quantities for mass, radius, metallicity and C/O.} Posteriors for models with and without a temperature inversion are shown in purple and dark cyan respectively.} 
\label{fig:w2220corner}
\end{figure}

\begin{figure}[h]
\centering
\includegraphics[width=.8\textwidth]{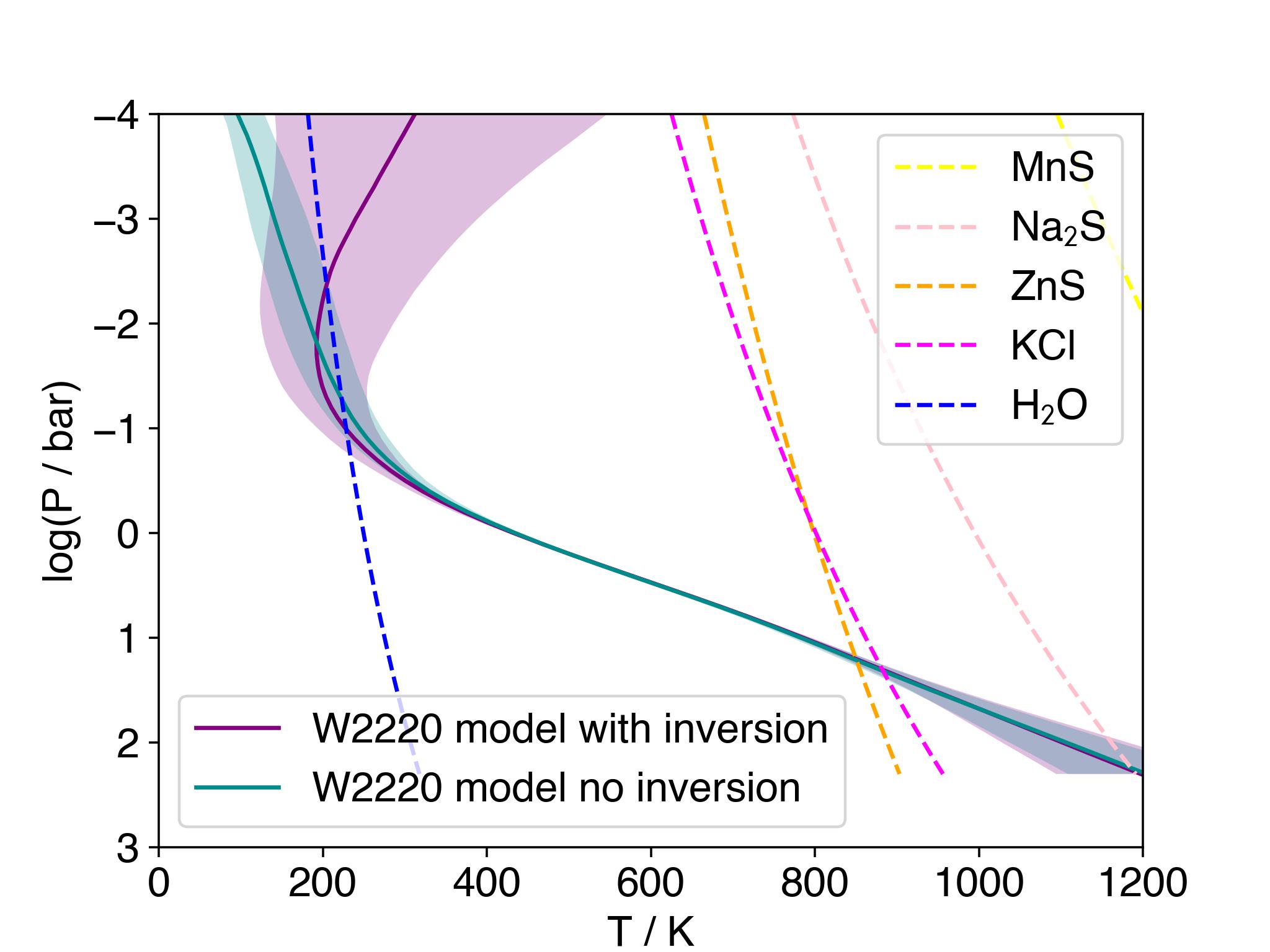} 
\caption{{\bf The retrieved thermal profiles for W2220 with and without an inversion.} Median posterior T/P profiles for the best fit model without an inversion, and the lower ranked model that allowed an inversion are shown in dark cyan and purple respectively.  The 67\% confidence intervals in model posteriors are indicated with shading for each. Also plotted with dashed lines are the condensation curves (assuming Solar composition gas) for possible cloud species.} 
\label{fig:w2220prof}
\end{figure}

\begin{figure}[h]
\centering
\includegraphics[width=1\textwidth]{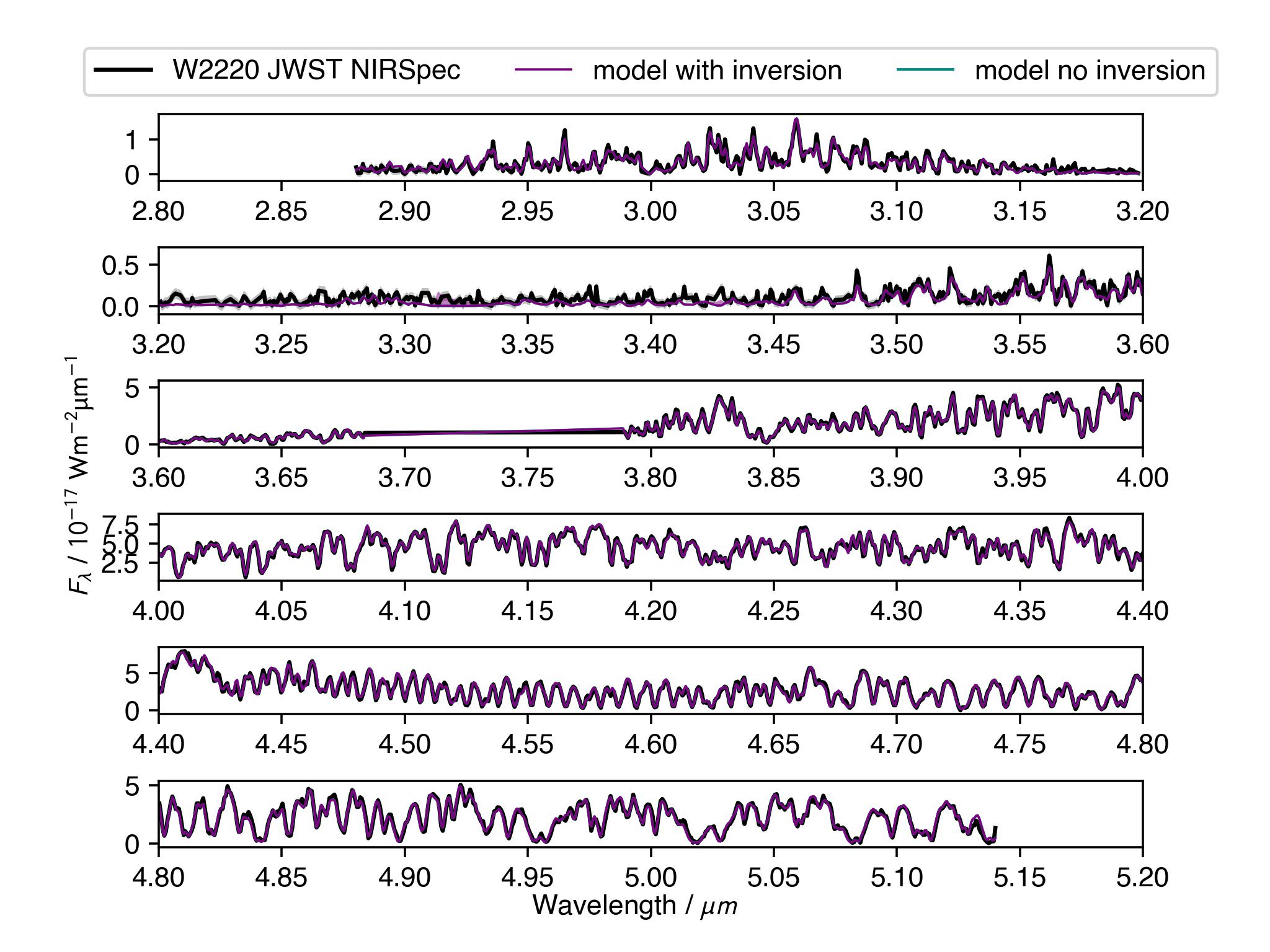} 
\caption{{\bf The JWST G395H spectrum for W2220.} Overlaid in purple and dark cyan are the median retrieval models for the source with and without temperature inversions. Data uncertainties are shaded in grey. The 67\% confidence intervals in the model posteriors are shaded in dark cyan and purple, but are of comparable extent to the line width.}
\label{fig:w2220spec}
\end{figure}

\begin{figure}[h]
\centering
\includegraphics[width=1.0\textwidth]{W1935_corner_comp-Extended-Data-Figure5.jpeg} 
\caption{{\bf The retrieved composition of W1935, along with gravity, and derived quantities for mass, radius, metallicity and C/O.}  Posteriors for models with and without a temperature inversion are shown in purple and dark cyan respectively.} 
\label{fig:w1935corner}
\end{figure}

\begin{figure}[h]
\centering
\includegraphics[width=1.0\textwidth]{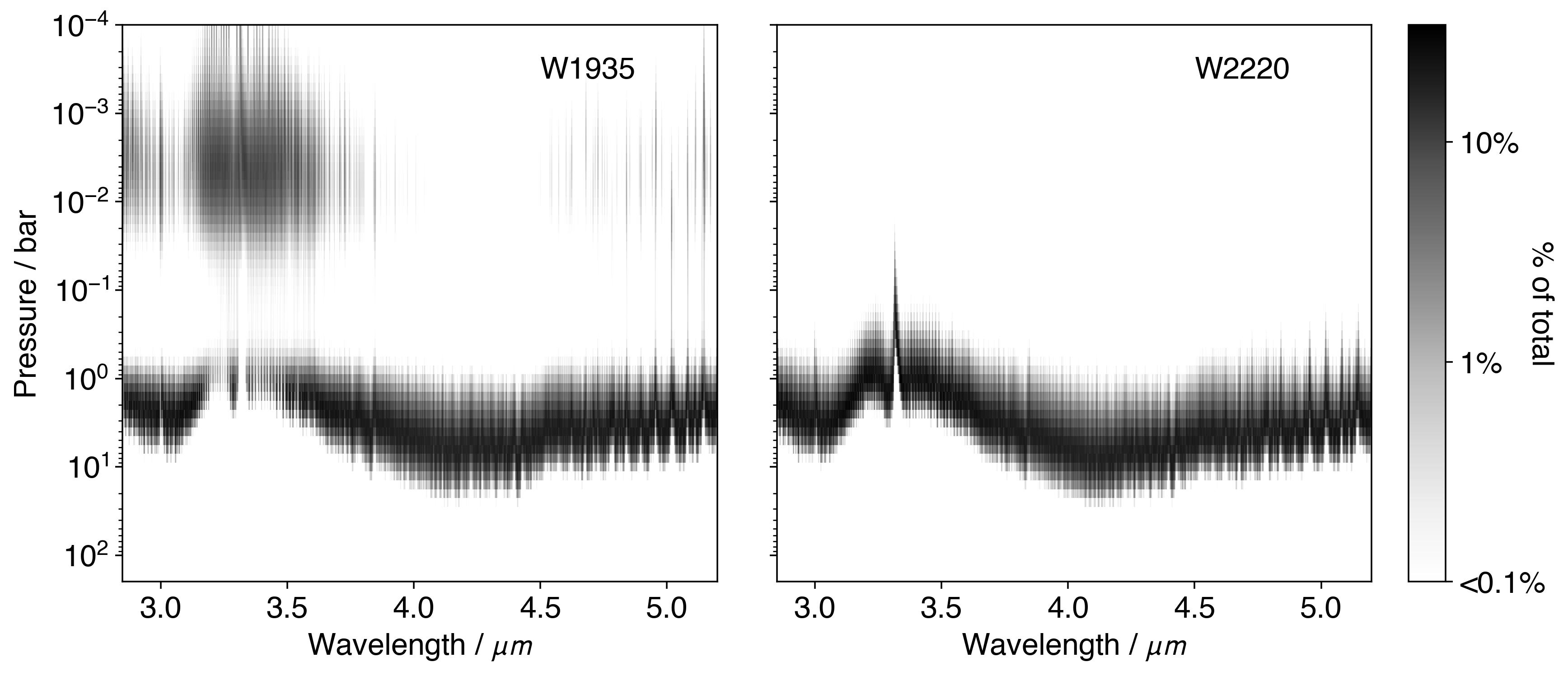} 
\caption{
{\bf The contribution functions for W1935 and W2220 maximum likelihood retrieval models.} The \chhhh\ emission seen in W1935 originates from pressures shallower than 0.01~bar.
} 
\label{fig:contrib}
\end{figure}

In addition to the free-parameters described above, we hardcode the following into our model atmosphere. 
In addition to the absorbing gases, we assume that the atmosphere is dominated by ${\rm H_2}$\ gas, with He present at a Solar H/He ratio. 
The total abundance of ${\rm H_2}$/He is set by the remaining fraction after the absorbing gas fractions have been accounted for. 

We calculate layer optical depths due to the absorbing gases using opacities sampled at a resolving power $R = 30,000$ taken from the compendium of \cite{freedman2008,freedman2014}, with updated opacities described in \cite{Burningham17} using the same method as \cite{burningham2021}. 
We generated the latest version of \chhhh\ line list \cite{hargreaves2020} with broadening coefficients relevant to $H_2$/He atmospheres using the computational methodology by \cite{GharibNezhad2021Opacity}, 
 and have incorporated \phhh\ opacities from \cite{marley2021,lupu2022}. 
We also include continuum opacities for H$_{2}$-H$_{2}$ and H$_{2}$-He collisionally induced absorption, using the cross-sections from  \cite{richard2012} and \cite{saumon2012}, and Rayleigh scattering due to H$_{2}$, He and CH$_{4}$ but neglect the remaining gases.
We also include continuum opacities due to bound-free and free-free absorption by H$^-$ \cite{john1988, Bell+1987}, and free-free absorption by H$_2^-$ \cite{bell1980}.

The emergent spectrum is convolved with a Gaussian kernel of width 2.2 pixels to simulate the instrumental profile of the NIRSpec G395H.   

In this work we coupled the forward model to the {\sc emcee} sampler \cite{emcee}, which is the same method as used in \cite{Burningham17,Gonzales18,Gonzales20,burningham2021,Vos22}. 
We use a log-likelihood function to assess the fit of the data to the model:

\begin{equation}
\ln \likL(\mathbf{y} \vert \mathbf{x}) = - \frac{1}{2}\sum_{i = 1}^{n} \frac{(y_{i} - F_{i}(\mathbf{x}))^2}{s^{2}_{i}} - \frac{1}{2} \ln(2 \pi s^{2}_{i})
\label{eqn:lnlik}
\end{equation}

where the index $i$ refers to the $i$th of $n$ spectral flux points, $y_i$, with errors $s_i$,  which are compared to the forward model fluxes $F_i$ for the current parameter set ${\bf x}$.
To (at least partially) account for model and other unaccounted sources of uncertainties we inflate our errors using a tolerance parameter, such that our data error, $s_i$, is given by: 

\begin{equation}
s_{i}^{2} = \sigma_{i}^{2} + 10^{b}
\end{equation}
where $\sigma_i$ is the measured error for the $i$th flux measurement, and $b$ is our tolerance parameter, which is retrieved \cite{emcee,hogg2010,tremaine2002}. 
The full set of parameters used and their priors are listed in Extended Data Table~\ref{tab:priors}. 

\begin{table}
\begin{tabular}{cc}
\hline
Parameter & Prior \\
\hline
gas fraction, ($X_{gas}$) & log-uniform, $\log X_{gas} \geq -12.0$, $\sum_{gas}{X_{gas}} \leq 1.0$ \\
thermal profile, $T$ & uniform, constrained by $0.0~{\rm K} < T_{i} < 5000.0~{\rm K}$ \\
profile smoothing parameter, $\gamma$ & uniform, $0 < \gamma < 10^5$ \\
gravity, $\log g$ & uniform, constrained by $1 \Mjup\  \leq gR^{2} / G \leq 80 \Mjup $ \\ 
scale factor, $R^{2} / D^{2}$ & uniform, constrained by $0.5 \Rjup\ \leq R \leq 2.0 \Rjup$ \\
rotational velocity, $v \sin i$ & uniform, $0\ {\rm km s^{-1}} < v\sin i < 100\ {\rm km s^{-1}}$ \\
radial velocity, $v_{rad}$ & uniform, $-250\ {\rm km s^{-1}} < v_{rad} < 250\ {\rm km s^{-1}}$  \\
tolerance factor, $b$ & uniform, $\log (0.01 \times min(\sigma_{i}^{2})) \leq b \leq \log(100 \times max(\sigma_{i}^{2}))$ \\
\hline
\end{tabular}
\caption{Parameters and priors adopted for the retrieval analysis. \label{tab:priors} }
\end{table}

We used 16 walkers per dimension. Following 10K iterations of burn-in, we ran our chains for blocks of 30K iterations, checking each time for convergence. 
In all cases convergence appeared to be achieved after the first 30K block, but we ran an additional 30K in each case nonetheless.

\subsubsection{Retrieval results}

We ran three models each for W1935 and W2220: our baseline model as described earlier; the baseline model adjusted with an additional prior that $T_i < T_{i+1}$, i.e. excludes the possibility of a temperature inversion; the baseline model adjusted to include \hhhplus\ as an absorbing gas at pressures less than 1 millibar. 
To distinguish the preferred model given our data we calculated the Bayesian Information Criterion (BIC) for each case. 
For a set of models, the one with the smallest (typically most negative) BIC will be the preferred model with the strength of the preference dependent on the value of the difference in the BICs, $\Delta BIC$. 
By convention, the ``winning'' model is defined as having $\Delta BIC =0 $, and lower-ranked models have $\Delta BIC$ quoted relative to that. 

\cite{kass1995} provide the following intervals for selecting between two models under the BIC:

\begin{itemize}
  \item $0  < \Delta BIC < 2$: no preference worth mentioning;
  \item $2  < \Delta BIC < 6$: positive;
  \item $6  < \Delta BIC < 10$: strong;
 \item $10  < \Delta BIC$: very strong.
\end{itemize}

We find that for W1935 the baseline model is very strongly preferred over the model that includes \hhhplus. Both of these models are very strongly preferred over the model that rules out a temperature inversion. 
By contrast, for W2220 we find that there is no preference worth mentioning between models with and without a temperature inversion, which are both strongly preferred over the model that includes \hhhplus.  This difference arises because W1935 shows emission in the \chhhh\ q-branch at 3.326~$\mu$m, whereas W2220 does not. 
The $\Delta BIC$ values for each model are given in Extended Data Table~\ref{tab:bic}.

\begin{table}
\begin{tabular}{l c}
\hline 
Model & $\Delta BIC$ \\
\hline
{\bf W1935} & \\
Baseline (allowed inversion) & 0 \\
No inversion & 63.6 \\
Baseline With \hhhplus & 9.9 \\
\hline
{\bf W2220}& \\
Baseline (allowed inversion) & 1.8 \\
No inversion & 0.0 \\
Baseline With \hhhplus & 9.9 \\
\hline\end{tabular}
\caption{$\Delta BIC$ values for the models tested for each target. 
\label{tab:bic}
}
\end{table}

We note that the inclusion of \hhhplus\ in the model had no impact on the quality of the fit, the maximum likelihood or the values of the other parameters. 
The abundance of \hhhplus\ was an upper limit of $\log f_{{\rm H_3^+}}\sim -6$.
We thus conclude that \hhhplus\ is not detected in either source. 
 
In the case of W2220, parameter estimates (not related to the temperature structure) are indistinguishable between the models that allow for a temperature inversion and those that do not.
A comparison of the posterior distributions for models with and without a temperature inversion is shown in a corner plot in Extended Data Figure~\ref{fig:w2220corner}.

The thermal profiles are also extremely similar, with only non-significant differences between the two. 
A comparison of the two retrieved thermal profiles is shown in Extended Data Figure~\ref{fig:w2220prof}.
It can be seen that the two profiles are identical at pressures deeper than $\log {\rm (P / bar)} = -2.0$, and do not deviate from one another significantly at shallower pressures.  
Extended Data Figure~\ref{fig:w2220spec} shows that the retrieved model spectra are also indistinguishable, and a similarly good fit to the data as the baseline model for W1935 (see  Figure~\ref{fig:w1935spec}).

In the case of W1935, the posteriors for the non-temperature related parameters are also very similar between the two models, as shown in Extended Data Figure~\ref{fig:w1935corner}. 
Although there are no significant differences between the compositions, there is a marginal trend across all the absorbing gases towards higher abundances in the baseline model (i.e. that allows for a temperature inversion).

Figures~\ref{fig:w1935spec} and~\ref{fig:retrievedprofile} demonstrate clearly that the model that allows a temperature inversion is the only one that is able to fit the \chhhh\ emission feature at 3.326~$\mu$m, which is absent in the spectrum of W2220.

Our retrieval results for both objects show some differences from those inferred from our SED-based luminosity and evolutionary model predictions (see Figures ~\ref{fig:w1935corner} and ~\ref{fig:w2220corner} as well as Extended Data Table~\ref{tab:target}). In both cases, the masses implied by our retrievals are slightly higher than the 1$\sigma$ upper limit suggested by evolutionary models. 
In the case of W1935 the difference is 1.3$\sigma$ and for W2220 it is 1.1$\sigma$.
It is not unusual for retrieval analyses to disagree with evolutionary models' predictions of $\log g$, radius, and mass, particularly in the absence of any strong prior evidence to provide empirical constraints \citep[e.g.][]{Burningham17,burningham2021,Gaarn2022,whiteford2023}. 
In addition, these retrievals cover a relatively narrow wavelength range which lacks recognized gravity sensitive spectral features. 
Establishing the presence and nature of any possible biases to higher gravity and mass estimates is beyond the scope of this work, and does not impact our central results or conclusions.

In Extended Data Figure~\ref{fig:contrib} we compare the contribution functions for maximum likelihood retrieval models for W1935 and W2220. 
We define the contribution function in an atmospheric layer, lying between pressures $P_1$ and $P_2$ is defined as:
\begin{equation}
C(\lambda, P) =\frac{B(\lambda,T(P))\int_{P_1}^{P_2}d\tau}{\exp{\int_0^{P_2}d\tau}}
\end{equation}

$C(\lambda, P)$ effectively maps the depth in the atmosphere from which the flux observed at a given wavelength originates. 
Extended Data Figure~\ref{fig:contrib} demonstrates that the \chhhh\ emission seen in W1935 originates from pressures shallower than 0.01~bar, where our model finds gas some 300~K hotter than is retrieved when no inversion is permitted.

For simplicity, we have only used cloud-free retrieval models. 
It has been well documented that the omission of clouds in a retrieval of an atmosphere which contains clouds can bias the thermal profile to a more isothermal gradient as the retrieval mimics the effect of clouds in screening hotter deeper layers in the opacity windows between molecular absorption features \cite[e.g. ][]{Burningham17,burningham2021,whiteford2023}. 
Hence, the kinks seen in the retrieved thermal profiles may be suggestive that clouds should be considered for future retrieval studies in this temperature regime.

\newpage

\backmatter
\bmhead{Availability of data and materials} The JWST data in this paper are part of GO program 2124 (PI J. Faherty) and are publically available in the Barbara A. Mikulski Archive for Space Telescopes (MAST; https://archive.stsci.edu/) under that program ID.  The HST WFC3 spectrum of W2220 is available from: https://ui.adsabs.harvard.edu/abs/2015ApJ...804...92S/abstract. 

\bmhead{Code availability} The data reduction pipeline jwst can be found at https://jwst-pipeline.readthedocs.io/en/latest/. The {\it Brewster} code is opensource and available at the following Github repository: https://github.com/substellar/brewster.  Similarly the SEDkit code is opensource and available at https://github.com/hover2pi/sedkit.  The setup which yields the results presented here-in and discussed in the methods section.

\bmhead{Acknowledgments}
JF acknowledges the Heising Simons Foundation as well as NSF award \#2009177, \#1909776, and NASA Award \#80NSSC22K0142.
BB acknowledges support from UK
Research and Innovation Science and Technology Facilities Council [ST/X001091/1]. J. M. V. acknowledges support from a Royal Society - Science Foundation Ireland University Research Fellowship (URF$\backslash$1$\backslash$221932).  Portions of this research were carried out at the Jet Propulsion Laboratory, California In- stitute of Technology, under a contract with the National Aeronautics and Space Administration.

\bmhead{Author contributions}
JF  oversaw all work including data reduction, analysis and modeling.  BB completed the atmospheric retrieval.  JG extracted the radial velocity of the source while GS, JV, SA, and RK contributed to the SED analysis.  CM, BL, MR, and EN contributed to the modeling analysis and DC, JDK, AM, AS, MK, DBG, CB, PE, CG, PE, CG, EG, FM, AR, and NW were all part of the original JWST GO program 2124 proposal which led to this data.

\bmhead{Author Information} The authors declare no competing interests.  Supplementary Information is available for this paper. Correspondence and requests for materials should be addressed to JF (jfaherty@amnh.org).

\clearpage

\clearpage


\end{document}